\documentclass[aps,pra,twocolumn,amsmath,amssymb,groupedaddress,nobibnotes,showpacs]{revtex4}

\usepackage{graphicx}
\usepackage{dcolumn}
\usepackage{bm}
\usepackage{amsmath}
\usepackage{amssymb,amsthm}
\usepackage{epsfig,color}
\usepackage[sans]{dsfont}

\definecolor{nblue}{rgb}{0.2,0.2,0.7}
\definecolor{ngreen}{rgb}{0.2,0.6,0.2}
\definecolor{nred}{rgb}{0.7,0.2,0.2}
\definecolor{nblack}{rgb}{0,0,0}

\def\N{\mathcal{N}}
\newcommand{\I}{\mathrm{I}}
\newcommand{\II}{\mathrm{II}}

\def\bea{\begin{eqnarray}}
\def\eea{\end{eqnarray}}

\begin{document}

\title{Entanglement behavior of quantum states of fermionic system in accelerated frame}

\author{Jinho Chang}

\author{Younghun Kwon}
\email{yyhkwon@hanyang.ac.kr}

\affiliation{Department of Physics, Hanyang University, Ansan,
Kyunggi-Do, 425-791, South Korea }

\date{\today}

\begin{abstract}

In this paper, we investigate the behavior of bipartite entanglement
of fermonic systems when one of parties is traveling with a uniform
acceleration. For the ordering problem in fermonic systems, we apply
the recent result in [Montero and Mart\'{i}n-Mart\'{i}nez, Phys.
Rev. A 83 052306 (2011)]. Based on the approach, we consider both
pure and mixed entangled states, and we show that the behavior in
terms of the entanglement measure, negativity, allows one to obtain
physical results, i.e. its convergence in the infinite acceleration.
The behavior shows that the ordering employed is relevant to derive
physical results for fermonic entanglement. This also corrects the
previous analysis of [Mart\'{i}n-Mart\'{i}nez and Fuentez.

\end{abstract}

\maketitle

\section{Introduction}
\label{intro}

Entanglement is one of central characteristics in quantum
information theory, as it contains correlations that do not have a
classical counterpart. Entangled states can be prepared by
correlations stronger than those of local classical systems assisted
with classical communication, but they do not exhibit arbitrarily
strong correlations. Therefore, quantum theory is consistent with
relativistic theories.

Along the line, it is of fundamental question in relativistic
quantum information theory how the behavior of entanglement can be
described in a relativistic setting, particularly when one of
parties is traveling in an acceleration \cite{ref:alsing1}
\cite{ref:ball}. This can be seen from two parties who share
entangled states, while one remains in an inertial frame and the
other is described in accelerated frame. It is highly nontrivial
that, even though entanglement in bosonic systems disappears in the
limit of infinite acceleration, entanglement in fermionic systems
can survive in the limit \cite{ref:fuentes}\cite{ref:alsing2}. This
is remarkable in that entanglement related to correlations can
behave differently, depending on fields, and can also persist in
such a limit. However the analysis for the fermionic system was only
based on the single-mode
approximation\cite{ref:fuentes}\cite{ref:alsing2}. To obtain a
precise understanding in the limit, it follows that one has to give
a full consideration beyond single-mode approximation. This has been
attempted, for instance, in Ref. \cite{ref:bruschi}, which is
however, hard to interpret physically. The question in fact lies
with the peculiarity of fermionic system, that gives rise to the
ambiguity in the ordering of operators. Note that, as it is
mentioned in Ref. \cite{ref:montero1}, what matters is to find an
ordering that has the physical relevance, i.e. such that
entanglement is characterized by what is observed by detectors. In
Ref. \cite{ref:montero1}, such an ordering beyond single mode
approximation is also proposed. In ref.\cite{ref:montero2} the
ordering is then applied to explain entanglement in the infinite
acceleration so the physicality of the fermionic structure is
provided and the independence of the choice of Unruh mode in the
infinite acceleration limit is discussed.\\
 However the ordering was tested only in the example
considered in Ref. \cite{ref:montero2}. Therefore in this work,
beyond the single-mode approximation, we investigate the behavior of
entanglement of fermonic systems using the recent construction
proposed in Ref. \cite{ref:montero1}, for both pure and mixed
states. In this way, the construction is extensively tested for a
number of entangled states. We show that in all of these cases, the
ordering constructed leads to convergence of fermonic entanglement
in the infinite acceleration, i.e. it yields physical results. This
also corrects the previously known analysis on the entanglement
behavior in Ref. \cite{ref:martin}. Finally, our results provide
strong evidence that the ordering suggested recently would
characterize entanglement as it gives physical results.

The organization of this article is as follows. In Sec. \ref{acc},
we will briefly review how fermionic systems are described in a
non-inertial frame. In Sec. \ref{ex}, the approach in Ref.
\cite{ref:montero1} is applied and the entanglement behavior of the
fermionic system in an accelerated frame is shown. In Sec. \ref{con}
we will conclude and discuss our results.

\section{ Accelerated Frame}
\label{acc}

\begin{figure}
\includegraphics[width=9.5cm]{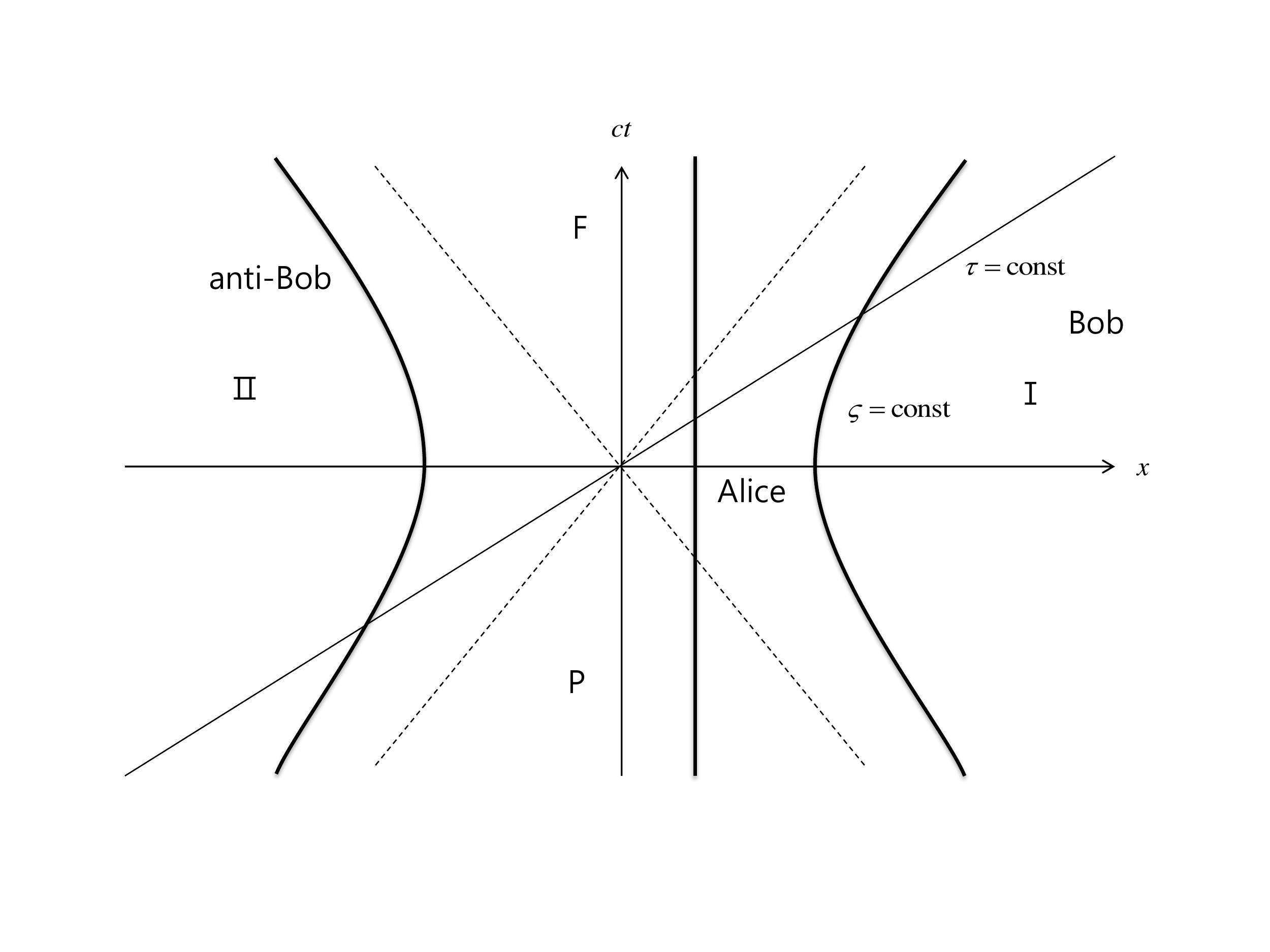}
\caption{The Rindler diagram is shown, where Alice is in the
inertial frame while Bob is in the uniformly accelerated frame. Two
regions $\I$ and $\II$ are causally disconnected. Transformations of
coordinates among regions $\I$, $\II$, $F$, and $P$ are explained in
the text. }
\label{fig1}
\end{figure}

 Let us begin with quantum fields in relativistic frames. A party
traveling with a uniform acceleration is described by the so-called
Rindler coordinate  $(\tau,\varsigma,y,z)$, which has the following
relation with Minkowski coordinate $(t,x,y,z)$,
\begin{equation}
ct=\varsigma \sinh(\frac{a \tau}{c}),x=\varsigma \cosh(\frac{a \tau}{c}) \label{eq1},
\end{equation}
where $a$ is a fixed acceleration of the frame and $c$ is the
velocity of light. For fixed $\varsigma$, the coordinate can be
found as hyperbolic trajectories in space-time. This is shown in
Fig. \ref{fig1}. Equation (\ref{eq1}) only covers the region (I) in
Fig. \ref{fig1}. The region (II) is covered by $ ct=-\varsigma
\sinh(\frac{a \tau}{c}),x=-\varsigma \cosh(\frac{a \tau}{c})$. The
other two regions (F and P) can be described as $ ct=\pm\xi
\cosh(\frac{a \sigma}{c}),x=\pm\xi \sinh(\frac{a \sigma}{c})$.

 Now let us consider the field in Minkowski and Rindler spacetime, which can be written as
\begin{eqnarray}
\phi &=& N_{M}\sum_{i}(a_{i,M}v^{+}_{i,M} +
b^{\dag}_{i,M}v^{-}_{i,M} )\nonumber\\
 &=& N_{R}\sum_{j}(a_{j,\I}v^{+}_{j,\I} +
b^{\dag}_{j,I}v^{-}_{j,\I} + a_{j,II}v^{+}_{j,\II} +
b^{\dag}_{j,II}v^{-}_{j,\II} ), \nonumber
\end{eqnarray}
where  $N_{M}$ and $N_{R}$ are normalization constants. Also
$v^{\pm}_{i,M}$ denotes the positive and negative energy solutions
of the Dirac equation in Minkowski spacetime, which can be obtained
with respect to the Killing vector field in Minkowski spacetime, and
$v^{\pm}_{i,\I}$ and $v^{\pm}_{i,\II}$ are the positive and negative
energy solutions of the Dirac equation in Rindler spacetime, with
respect to the Killing vector field in regions I and II. In addition
$a^{\dag}_{i,\Delta}(a_{i,\Delta})$ and
$b^{\dag}_{i,\Delta}(b_{i,\Delta})$ are the creation (annihilation)
operators for the positive and negative energy solutions (particle
and antiparticle), where $\Delta $ denotes $M,\I,\II$. Then they
satisfy the anticommutation relations
$\{a_{i,\Delta},a^{\dag}_{j,\Delta^{'}}\}=\{b_{i,\Delta},b^{\dag}_{j,\Delta^{'}}\}=\delta_{ij}\delta_{\Delta
\Delta^{'} }$. It is known that the combination of Minkowski mode,
called Unruh mode, can be transformed into a monochromatic Rindler
mode and can annihilate the same Minkowski vacuum. The following
relation thus holds:
\begin{equation}
A_{i,R/L}\equiv \cos \gamma_{i}a_{i,\I/ \II} - \sin \gamma_{i} b^{\dag}_{i,\II / \I}, \nonumber
\end{equation}
where $\cos \gamma_{i}=(e^{\frac{-2 \pi \Omega c}{a}}+1)^{-1/2}$. A more general relation can also be found,
\begin{equation} a^{\dag}_{i,U}=q_{L}(A^{\dag}_{\Omega ,L}\otimes
I_{R}) + q_{R}(I_{L} \otimes  A^{\dag}_{\Omega ,R}), \label{eq2}
\end{equation}
by which one can go beyond the single-mode approximation.

Using these relations, in case of Grassmann scalar, the Unruh vacuum can be given by
\begin{eqnarray}
|0_{\Omega }\rangle_{U} &=& \cos^{2} \gamma_{\Omega }
|0000\rangle_{\Omega } - \sin \gamma_{\Omega } \cos \gamma_{\Omega }
|0011\rangle_{\Omega }
\nonumber\\
        &+& \sin \gamma_{\Omega } \cos \gamma_{\Omega } |1100\rangle_{\Omega } - \sin^{2}
\gamma_{\Omega } |1111\rangle_{\Omega }\label{eq3}
\end{eqnarray}
Here we use the notation $|pqmn\rangle_{\Omega } \equiv |p_{\Omega
}\rangle^{+}_{\I}|q_{\Omega }\rangle^{-}_{\II}  | m_{\Omega
}\rangle^{-}_{\I} |n_{\Omega }\rangle^{+}_{\II} $. Then,
one-particle states can be obtained as\begin{eqnarray} |1_{\Omega
}\rangle^{+}_{U} &=& q_{R}(\cos \gamma_{\Omega }
|1000\rangle_{\Omega } - \sin \gamma_{\Omega }
|1011\rangle_{\Omega})
\nonumber\\
                        &+& q_{L}(\sin \gamma_{\Omega } |1101\rangle_{\Omega } + \cos
 \gamma_{\Omega } |0001\rangle_{\Omega }), \label{eq4}
 \eea
and,
\bea
|1_{\Omega }\rangle^{-}_{U} &=& q_{L}(\cos \gamma_{\Omega }
|0100\rangle_{\Omega } - \sin \gamma_{\Omega } |0111\rangle_{\Omega})
\nonumber\\
                        &+& q_{R}(\sin \gamma_{\Omega } |1110\rangle_{\Omega } + \cos
 \gamma_{\Omega } |0010\rangle_{\Omega }). \label{eq5}
\end{eqnarray}
From now on, for both convenience and simplicity, we restrict our
consideration to cases when $q_{R}$ and $q_{L}$ are real numbers,
and we also omit the index $\Omega$ throughout. In Eqs. (\ref{eq4})
and (\ref{eq5}), the single-mode approximation can be found by
putting $q_{R}=1$, so that the vacuum in the Minkowski frame can be
written as, $|0\rangle_{M} = \cos \gamma
|0\rangle_{\I}|0\rangle_{\II} + \sin \gamma
|1\rangle_{\I}|1\rangle_{\II}$.

Very recently, in Refs. \cite{ref:montero1} \cite{ref:montero2} it
has been suggested that the ordering in fermonic systems should be
rearranged by the sequence of particles and antiparticles in the
separated regions, so that the entanglement behavior of those states
will yield physical results. In the following section, we apply the
construction and derive the entanglement behavior of fermonic
systems.

\section{Entanglement behavior}
\label{ex}

In this section, we consider the construction shown in Refs.
\cite{ref:montero1} \cite{ref:montero2}, and we derive the
entanglement behavior accordingly. For bipartite pure entanglement,
the entanglement property of quantum states can be simplified
greatly due to the Schmidt decomposition, by which a given quantum
state can be expressed using a single parameter. That is, the
decomposition allows any bipartite pure state to be in the
expression $|\psi (\alpha) \rangle = \cos\alpha |00\rangle +
\sin\alpha|11\rangle$ for some $\alpha$, with a set of orthonormal
basis $|0\rangle$ and $|1\rangle$.

To quantify entanglement,we apply the measure called
\emph{negativity}, which is based on the partial transpose of
quantum states, i.e. taking transpose to the state of either system
of two parties \cite{ref:peres}. Then, for a given quantum state
$\rho$, negativity $\N$ can be computed as follows: $\N(\rho) =
\sum_{i} | \lambda_i | $ in which $\lambda_i$ are negative
eigenvalues of $\rho^{\Gamma}$, where $\Gamma$ denotes the partial
transpose \cite{ref:neg}. Note that the measure-negativity-is useful
as a computable measure in various contexts.

With these, the entanglement behavior is to be studied in the
following scenario. Suppose that two parties, called Alice and Bob,
share entangled states in inertial frames in the beginning.
Afterward, Bob moves with a uniform acceleration. We then show that
entanglement in the infinite acceleration allows us to obtain
physical results. In particular, results shown in Sec. \ref{sub1},
\ref{sub2}, and \ref{sub3} correct previously known analysis in Ref.
\cite{ref:martin}.

\subsection{Bipartite pure states I - particle and antiparticle Unruh excitations}
\label{sub1}

We first consider pure entanglement between Alice and Bob. As was
mentioned, suppose that two parties share a pure state and Bob
travels with a uniform acceleration. Then, the state shared is
described as
\begin{equation}
|\Phi_{+} (\alpha) \rangle = \cos \alpha |0\rangle_{M}|0\rangle_{U} + \sin \alpha
|1\rangle_{M}|1^{+}\rangle_{U}. \label{eq6}
\end{equation}

Suppose that Bob's detector cannot distinguish between the particle
or the antiparticle. As it is depicted in Fig. \ref{fig1}, two
regions $\I$ and $\II$ are causally disconnected and thus Bob does
not have assess to both. Hence, the state between Alice and Bob can
be found by tracing either regions. First, the state of Alice and
Bob when Bob is in region $\I$ is described as
\begin{eqnarray}
\rho_{AB_{\mathrm{I}}}^{\Phi_{+}}
&=& \cos ^2 \alpha \cos ^4 \gamma |000\rangle \langle 000| \nonumber\\
&+& \frac{q_{R}}{2} \sin 2\alpha \cos ^{3} \gamma (|000\rangle \langle 110| + |110\rangle \langle 000|) \nonumber\\
&+& q_{L}^{2} sin ^{2} \alpha \cos ^{2} \gamma |100\rangle \langle 100| \nonumber\\
&+& \frac{1}{2}(1+(1-2 q_{L}^{2}) \cos 2\gamma) \sin ^{2}\alpha
|110\rangle \langle 110| \nonumber\\
&-&\frac{q_{L}}{2}  \sin 2\alpha \cos ^{2}\gamma \sin
\gamma(|001\rangle \langle 100| + |100\rangle
\langle 001|) \nonumber\\
&-& \frac{q_{R}q_{L}}{2} \sin ^{2}\alpha \sin 2\gamma (|100\rangle
\langle 111| + |111\rangle \langle 100| ) \nonumber\\
&+& \frac{1}{4} \cos ^{2}\alpha  \sin ^{2} 2\gamma (|001\rangle
\langle 001| + |010\rangle \langle 010|) \nonumber\\
&+& \frac{q_{R}}{2} \sin 2\alpha \cos \gamma \sin ^{2} \gamma
(|001\rangle \langle 111| + |111\rangle \langle 001|) \nonumber\\
&+& q_{R}^{2} \sin ^{2} \alpha \sin ^{2} \gamma |111\rangle \langle 111| \nonumber\\
&+& \frac{q_{L}}{2} \sin 2\alpha \sin ^{3} \gamma (|011\rangle \langle 110| + |110\rangle \langle 011|) \nonumber\\
&+& \cos ^{2} \alpha \sin ^{4} \gamma |011\rangle \langle 011|. \nonumber
\end{eqnarray}
And, the state of Alice and antiBob (i.e. in Bob's region II) is
then expressed after tracing the region $\I$, as follows:
\begin{eqnarray}
\rho_{AB_{\mathrm{II}}}^{\Phi_{+}}
&=& \cos ^2 \alpha \cos ^4 \gamma |000\rangle \langle 000| \nonumber\\
&+& \frac{q_{L}}{2} \sin 2\alpha \cos ^{3} \gamma (|000\rangle \langle 110| + |110\rangle \langle 000|) \nonumber\\
&+& q_{R}^{2} \sin ^{2} \alpha \cos ^{2} \gamma |100\rangle \langle 100| \nonumber\\
&+& \frac{1}{2}(1+(1-2 q_{L}^{2}) \cos 2\gamma) \sin ^{2}\alpha |110\rangle \langle 110| \nonumber\\
&+& \frac{q_{R}}{2}  \sin 2\alpha \cos ^{2}\gamma \sin \gamma(|001\rangle \langle 100| + |100\rangle \langle 001|) \nonumber\\
&-& \frac{q_{R}q_{L}}{2} \sin ^{2}\alpha \sin 2\gamma (|100\rangle \langle 111| + |111\rangle \langle 100| ) \nonumber\\
&+& \frac{1}{4} \cos ^{2}\alpha \sin ^{2} 2\gamma (|001\rangle \langle 001| + |010\rangle \langle 010|) \nonumber\\
&-& \frac{q_{L}}{2} \sin 2\alpha \cos \gamma sin ^{2} \gamma (|001\rangle \langle 111| + |111\rangle \langle 001|) \nonumber\\
&+& q_{L}^{2} \sin ^{2} \alpha \sin ^{2} \gamma |111\rangle \langle 111| \nonumber\\
&+& \frac{q_{R}}{2} \sin 2\alpha \sin ^{3} \gamma (|011\rangle \langle 110| + |110\rangle \langle 011|) \nonumber\\
&+& \cos ^{2} \alpha \sin ^{4} \gamma |011\rangle \langle 011|. \nonumber
\end{eqnarray}
Note that these are exact expressions, not obtained from the
single-mode approximation.

Entanglement of these states is estimated using negativity in Ref.
\cite{ref:neg}, which is shown in Fig. 2. As $\gamma$ increases, the
entanglement of $ \rho_{AB_{\mathrm{I}}}^{\Phi_{+}}$ decreases but
that of $ \rho_{AB_{\mathrm{II}}}^{\Phi_{+}}$ increases. They
eventually coincide at $\gamma=\frac{\pi}{4}$. It is thus shown that
for states $ \rho_{AB_{\mathrm{I}}}^{\Phi_{+}}$ and $
\rho_{AB_{\mathrm{II}}}^{\Phi_{+}}$, entanglement is determined
independently of $q_{R}$ at the infinite acceleration,
$\gamma=\frac{\pi}{4}$. As has been discussed in Ref.
\cite{ref:montero2}, these allow us to obtain physical results since
at infinite acceleration, entanglement of the states $
\rho_{AB_{\mathrm{I}}}^{\Phi_{+}}$ and $
\rho_{AB_{\mathrm{II}}}^{\Phi_{+}}$ is shown to be independent on
$q_{R}$.
\begin{figure}[h!]
\begin{center}
\includegraphics[width=8cm]{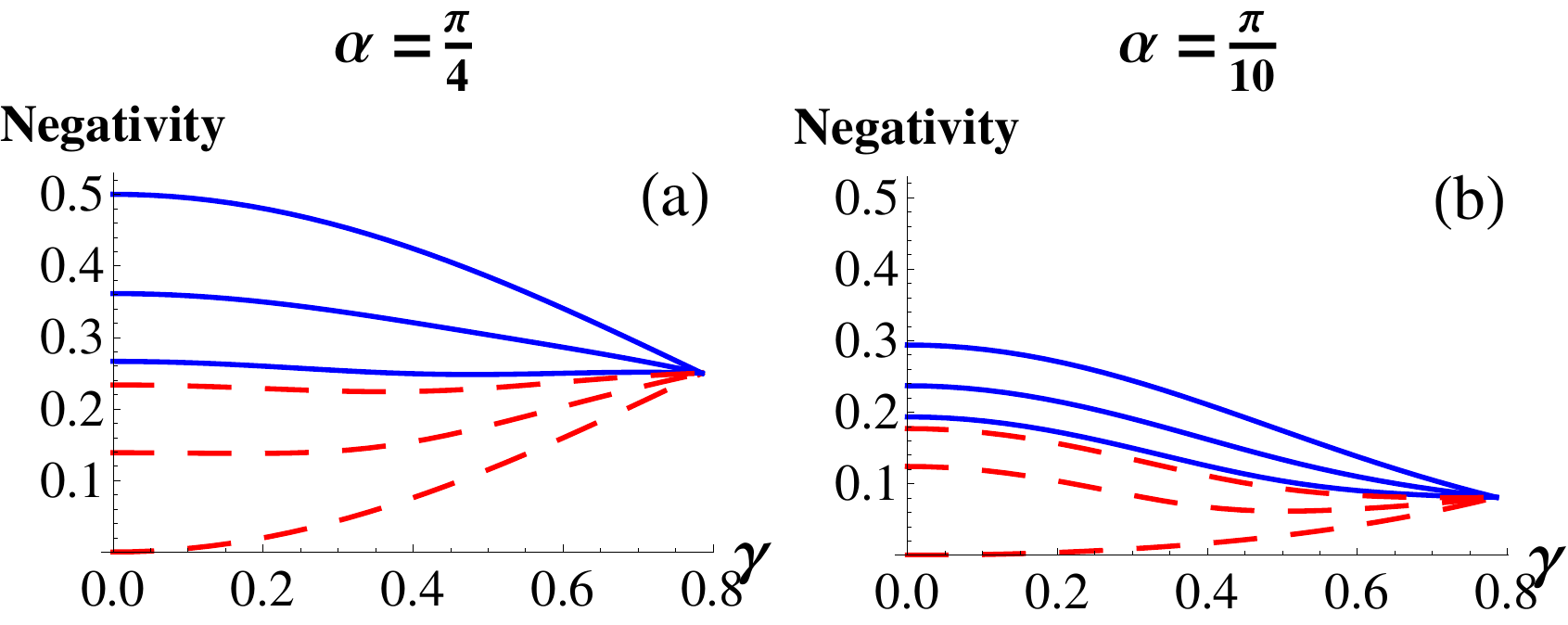}
\caption{\label{fig2}(Color online) Negativity is computed for
states $ \rho_{AB_{\mathrm{I}}}^{\Phi_{+}}$ and $
\rho_{AB_{\mathrm{II}}}^{\Phi_{+}}$. Parts (a) and (b) show the
cases of $\alpha=\frac{\pi}{4}$ and $\alpha=\frac{\pi}{18}$
respectively. The blue solid(the red dotted) lines from top to
bottom(from bottom to top) denote the negativity of $
\rho_{AB_{\mathrm{I}}}^{\Phi_{+}}$($
\rho_{AB_{\mathrm{II}}}^{\Phi_{+}}$) at $q_{R}=1$,$q_{R}=0.85$ and
$q_{R}=0.73$ respectively. $\gamma=\frac{\pi}{4}$ means the infinite
acceleration . As it can be seen, the entanglement of $
\rho_{AB_{\mathrm{I}}}^{\Phi_{+}}$ and $
\rho_{AB_{\mathrm{II}}}^{\Phi_{+}}$ coincides at infinite
acceleration. }
\end{center}
\end{figure}

Next let us consider the case in which Bob and anti-Bob detector can
distinguish between the particle and the antiparticle. Then for the
state in Eq. (\ref{eq6}), one can find density matrices
corresponding to cases of the Alice-Bob particle in region I, the
Alice-Bob antiparticle in region I, the Alice-anti-Bob particle in
region II, and the Alice-anti-Bob antiparticle in region II. From
the density matrices, entanglement can be computed using negativity
for those cases, see Fig. 3.


\begin{figure}[h!]
\begin{center}  
\includegraphics[width=8cm]{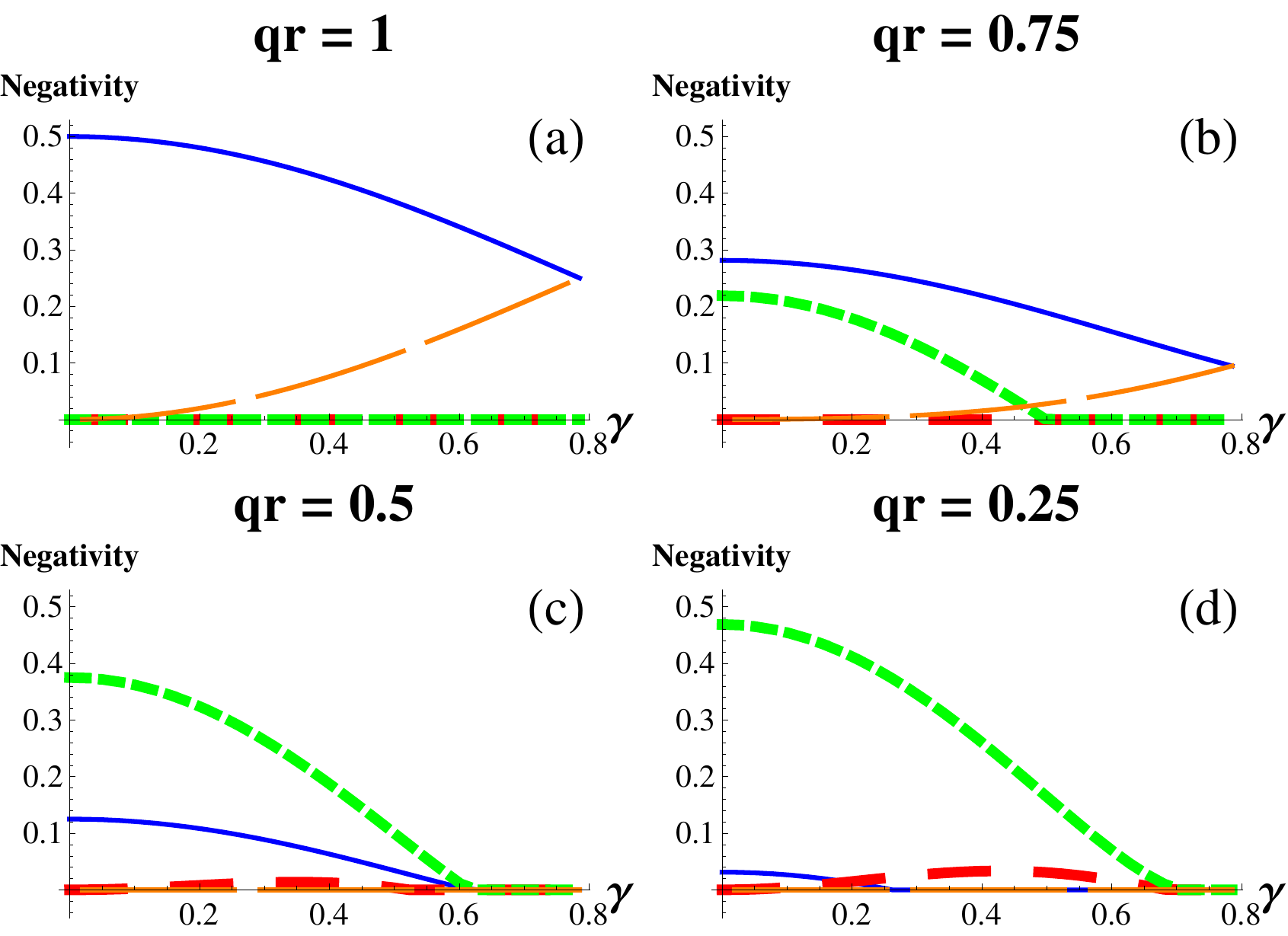}
\caption{\label{fig3}(Color online) For the state $\Phi_{+}$ in Eq.
(\ref{eq6}), negativity is computed for Alice-Bob particle in region
I, Alice-Bob antiparticle in region I, Alice-anti-Bob particle in
region II, and Alice-anti-Bob antiparticle in region II. Parts (a),
(b), (c) and (d) show the cases of
$q_{R}=1,q_{R}=0.75,q_{R}=0.5$,and $q_{R}=0.25$ respectively. The
blue solid line, the red thick dashed one, the green dotted one and
the orange dot-dashed one denote the negativity of Alice-Bob
particle in region I, Alice-Bob antiparticle in region I,
Alice-anti-Bob particle in region II, and Alice-anti-Bob
antiparticle in region II respectively. }
\end{center}
\end{figure}

 In Fig. 3 (a), the behavior of entanglement at $q_{R}=1$ in terms of
negativity is shown. As $\gamma$ increases, the entanglement of the
Alice-Bob particle in region I decreases, however that of the
Alice-anti-Bob antiparticle in region II increases. At
$q_{R}=\frac{1}{2}$, we can see nonzero valued negativity only in
cases of the Alice-Bob particle in region I, the Alice-Bob
antiparticle in region I and the Alice-anti-Bob particle in region
II.

\subsection{Bipartite pure states II - particle and antiparticle Unruh excitations}
\label{sub2}

We next consider the entanglement between Alice and Bob, when they
share the following state:
\begin{equation}
|\Phi_{-} (\alpha) \rangle = \cos \alpha |0\rangle_{M}|0\rangle_{U} + \sin \alpha |1\rangle_{M}|1^{-}\rangle_{U} \label{eq7},
\end{equation}
As it is explained previously, Bob has inaccessible part due to his
acceleration. The state of Alice and Bob after tracing the region
$\II$ is found as follows:
\begin{eqnarray}
\rho_{AB_{\mathrm{I}}}^{\Phi^{-}} &=& \cos ^2 \alpha \cos ^4 \gamma |000\rangle \langle 000| \nonumber\\
&+& \frac{q_{R}}{2} \sin 2\alpha cos ^{3} \gamma (|000\rangle \langle 101| + |101\rangle \langle 000|) \nonumber\\
&+& q_{L}^{2} \sin ^{2} \alpha \cos ^{2} \gamma |101\rangle \langle 101| \nonumber\\
&+& \frac{1}{2}(1+(1-2 q_{L}^{2}) \cos 2\gamma) \sin ^{2}\alpha
|101\rangle \langle 101| \nonumber\\
&+& \frac{q_{L}}{2}  \sin 2\alpha \cos^{2}\gamma sin
\gamma(|010\rangle \langle 100| + |100\rangle
\langle 010|) \nonumber\\
&-& \frac{q_{R}q_{L}}{2} \sin ^{2}\alpha \sin 2\gamma (|100\rangle
\langle 111| + |111\rangle \langle 100| ) \nonumber\\
&+& \frac{1}{4} \cos ^{2}\alpha  \sin ^{2} 2\gamma (|001\rangle
\langle 001| + |010\rangle \langle 010|) \nonumber\\
&-& \frac{q_{R}}{2} \sin 2\alpha \cos \gamma sin ^{2} \gamma
(|010\rangle \langle 111| + |111\rangle \langle 010|) \nonumber\\
&+& q_{R}^{2} \sin ^{2} \alpha \sin ^{2} \gamma |111\rangle \langle 111| \nonumber\\
&+& \frac{q_{L}}{2} \sin 2\alpha \sin ^{3} \gamma (|011\rangle \langle 101| + |101\rangle \langle 011|) \nonumber\\
&+& \cos ^{2} \alpha \sin ^{4} \gamma |011\rangle \langle 011|. \nonumber
\end{eqnarray}
The state that Alice-anti-Bob(in Bob's region II) share can be
obtained after tracing the region I,
\begin{eqnarray}
\rho_{AB_{\mathrm{II}}}^{\Phi^{-}} &=&  \cos ^2 \alpha \cos ^4 \gamma |000\rangle \langle 000| \nonumber\\
&+& \frac{q_{L}}{2} \sin 2\alpha \cos ^{3} \gamma (|000\rangle \langle 101| + |101\rangle \langle 000|) \nonumber\\
&+& q_{R}^{2} \sin ^{2} \alpha \cos ^{2} \gamma |100\rangle \langle 100| \nonumber\\
&+& \frac{1}{2}(1+(1-2 q_{R}^{2}) \cos 2\gamma) \sin ^{2}\alpha
|101\rangle \langle 101| \nonumber\\
&-&\frac{q_{L}}{2}  \sin 2\alpha \cos ^{2}\gamma \sin
\gamma(|010\rangle \langle 100| + |100\rangle
\langle 010|) \nonumber\\
&-& \frac{q_{R}q_{L}}{2} \sin ^{2}\alpha \sin 2\gamma (|100\rangle
\langle 111| + |111\rangle \langle 100| ) \nonumber\\
&+& \frac{1}{4} \cos ^{2}\alpha  \sin ^{2} 2\gamma (|001\rangle
\langle 001| + |010\rangle \langle 010|) \nonumber\\
&+& \frac{q_{R}}{2} \sin 2\alpha \cos \gamma \sin ^{2} \gamma
(|010\rangle \langle 111| + |111\rangle \langle 010|) \nonumber\\
&+& q_{L}^{2} \sin ^{2} \alpha \sin ^{2} \gamma |111\rangle \langle 111| \nonumber\\
&+& \frac{q_{L}}{2} \sin 2\alpha \sin ^{3} \gamma (|011\rangle \langle 101| + |101\rangle \langle 011|) \nonumber\\
&+& \cos ^{2} \alpha \sin ^{4} \gamma |011\rangle \langle 011|. \nonumber
\end{eqnarray}

Entanglement of states $ \rho_{AB_{\mathrm{I}}}^{\Phi^{-}}$ and
$\rho_{ AB_{\mathrm{II}}}^{\Phi^{-}}$ is shown in Fig. 4. Their
behavior are very similar to cases shown in Sec. \ref{sub1}. It is
also shown that entanglement of states $
\rho_{AB_{\mathrm{I}}}^{\Phi_{-}}$ and $
\rho_{AB_{\mathrm{II}}}^{\Phi_{-}}$ is independent of $q_{R}$ at
$\gamma=\frac{\pi}{4}$. This demonstrates proper entanglement
behavior as it yields physical results.

\begin{figure}[h!]
\begin{center}  
  \includegraphics[width=8cm]{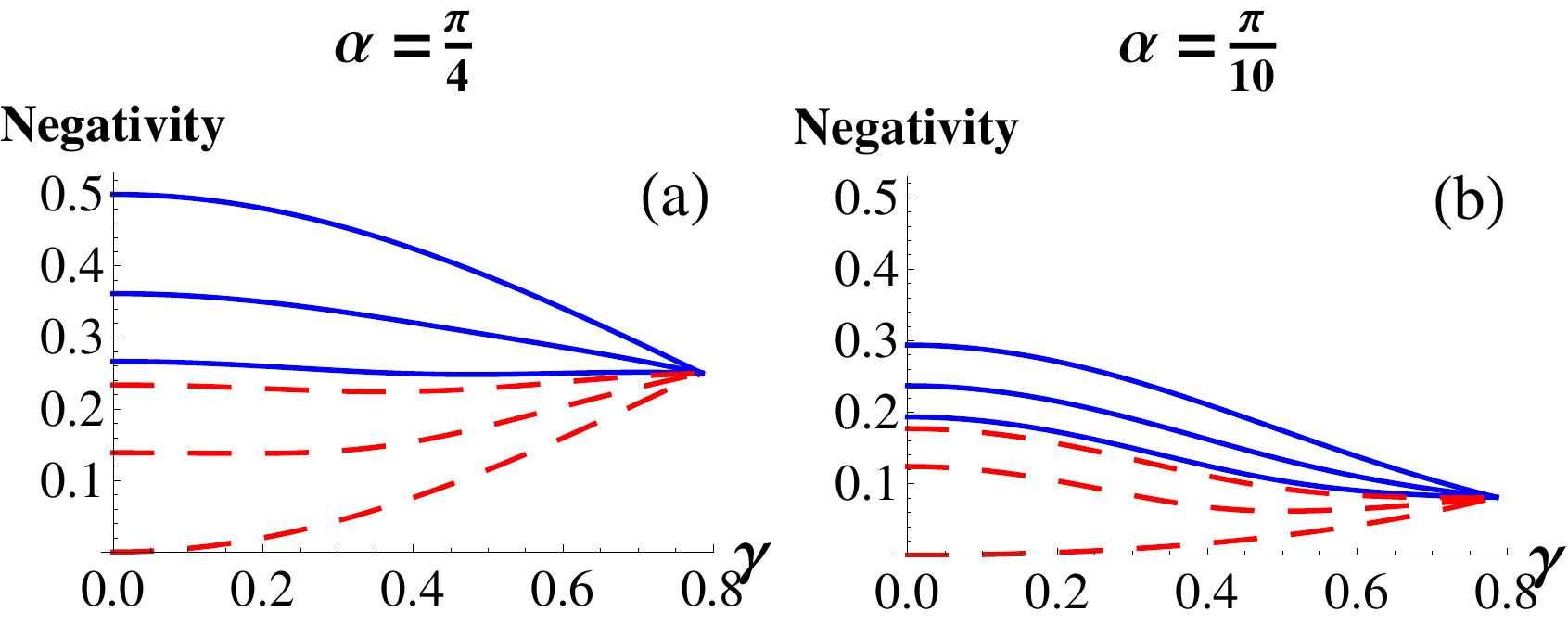}
  \caption{\label{fig4}(Color online) Negativity
 is computed for states $ \rho_{ AB_{\mathrm{I}}}^{\Phi^{-}}$ and $
\rho_{ AB_{\mathrm{II}}}^{\Phi^{-}}$. Parts (a) and (b) show the
cases of $\alpha=\frac{\pi}{4}$ and $\alpha=\frac{\pi}{18}$
respectively. The blue solid(the red dotted) lines from top to
bottom(from bottom to top) denote the negativity of $
\rho_{\Phi^{-}AB_{I}}$($ \rho_{\Phi^{-}AB_{II}}$) at
$q_{R}=1$,$q_{R}=0.85$ and $q_{R}=0.73$ respectively.
$\gamma=\frac{\pi}{4}$ means the infinite acceleration. As it is
shown, both states $ \rho_{AB_{\I}}^{\Phi_{-}}$ and $
\rho_{AB_{\II}}^{\Phi^{-}}$ have the same value in negativity in the
infinite acceleration. }
\end{center}
\end{figure}

\begin{figure}[h!]
\begin{center}  
  \includegraphics[width=8cm]{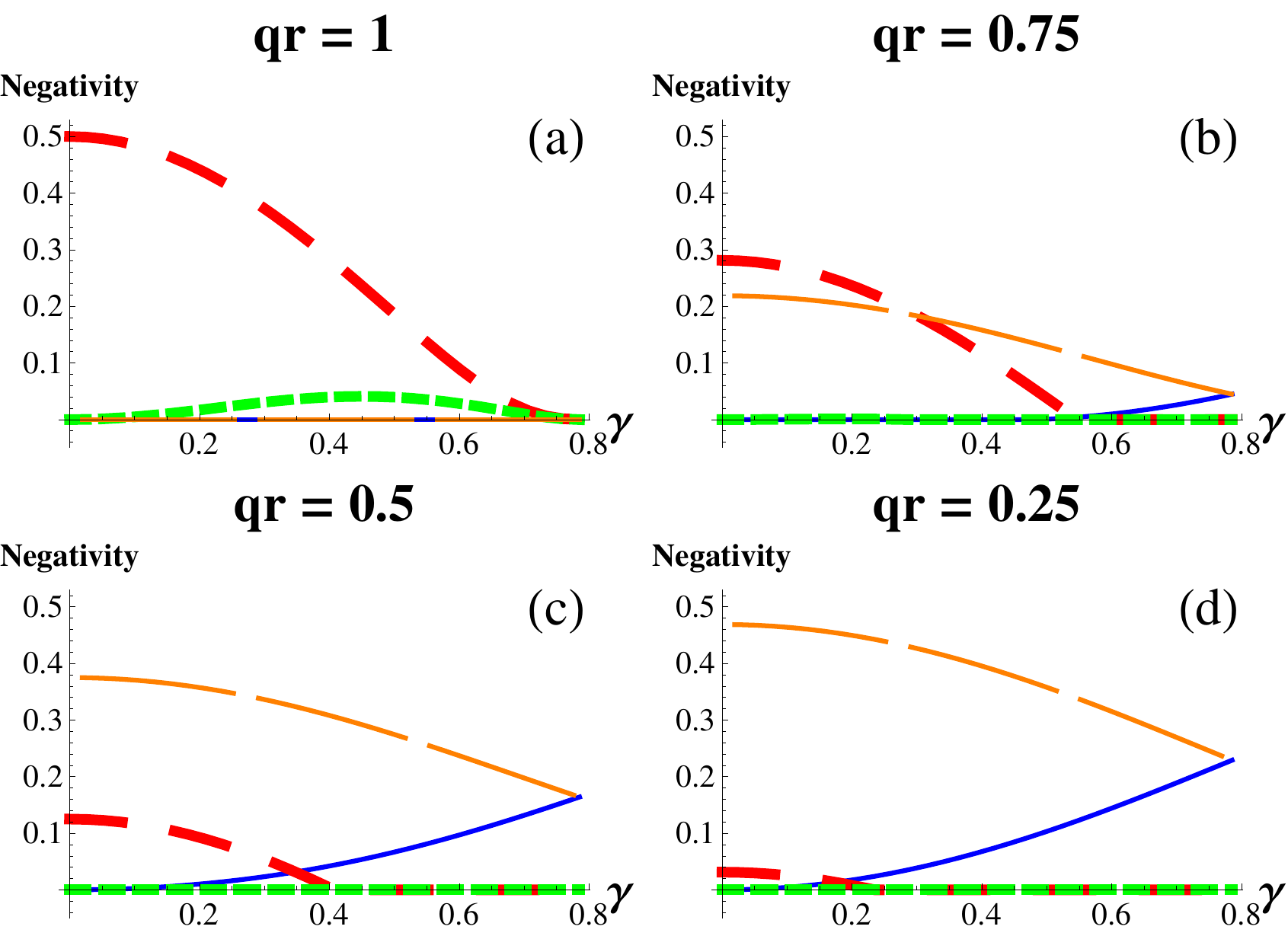}
  \caption{\label{fig5}(Color online) For
 the state $\Phi_{-}$ in Eq. (\ref{eq7}), negativity is computed for
Alice-Bob particle in region I, Alice-Bob antiparticle in region I,
Alice-anti-Bob particle in region II, and Alice-anti-Bob
antiparticle in region II. Parts (a),(b),(c) and (d) show the cases
of $q_{R}=1,q_{R}=0.75,q_{R}=0.5$,and $q_{R}=0.25$ respectively. The
blue solid line, the red thick dashed one, the green dotted one and
the orange dot-dashed one denote negativity of Alice-Bob particle in
region I, Alice-Bob antiparticle in region I, Alice-anti-Bob
particle in region II, and Alice-anti-Bob antiparticle in region II
respectively.}
\end{center}
\end{figure}

As we discussed in the previous section, if the Bob and anti-Bob
detector can distinguish between the particle and the antiparticle,
for the state $\Phi_{-}$ in Eq. (\ref{eq7}) one may find the density
matrices of the Alice-Bob particle in region I, the Alice-Bob
antiparticle in region I, the Alice-anti-Bob particle in region II,
and the Alice-anti-Bob antiparticle in region II. In this way,
density matrices corresponding to cases of the Alice-Bob particle in
region I, the Alice-Bob antiparticle in region I, the Alice-anti-Bob
particle in region II, and the Alice-anti-Bob antiparticle in region
II can be obtained. In Fig. \ref{fig5}, the entanglement behavior is
shown for different values of $q_{R}$.


 We observe that the entanglement behavior shown in Fig. 5 is
different from those in Fig. 3. At $q_{R}=1$ the non-zero values in
negativity can be found in cases of the Alice-Bob antiparticle in
region I and the Alice-anti-Bob particle in region II, but not in
the others. At $q_{R}=\frac{3}{4}$, non-vanishing entanglement is
found in cases of the Alice-Bob particle in region I, the Alice-Bob
antiparticle in region I, and the Alice-anti-Bob antiparticle in
region II.

\begin{figure}
\begin{center}  
  \includegraphics[width=8cm]{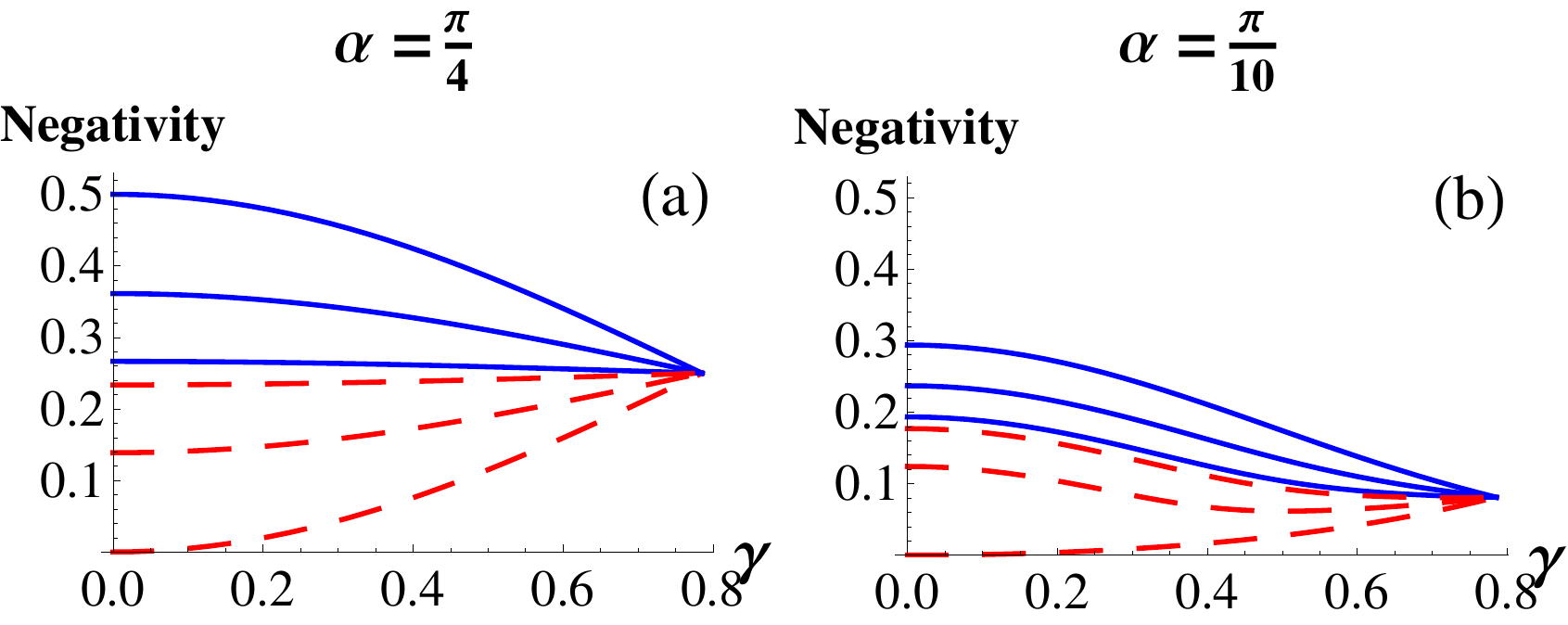}
  \caption{\label{fig6}(Color online) The entanglement behavior
is shown for cases considered in Sec. \ref{sub3}, two states $
\rho_{AB_{\mathrm{I}}}^{\Phi_{*}}$ and $ \rho_{
AB_{\mathrm{II}}}^{\Phi_{*}}$. Parts (a) and (b) show the behavior
when $\alpha=\frac{\pi}{4}$ and $\alpha=\frac{\pi}{18}$,
respectively. The blue solid(the red dotted) lines from top to
bottom(from bottom to top) denote the negativity of $
\rho_{AB_{\mathrm{I}}}^{\Phi_{*}}$ ($ \rho_{ AB_{\II}}^{\Phi_{*}}$)
at $q_{R}=1$,$q_{R}=0.85$ and $q_{R}=0.73$ respectively. Note that
the infinite acceleration is meant when $\gamma=\frac{\pi}{4}$. It
is shown that entanglement of two states $
\rho_{AB_{\I}}^{\Phi_{*}}$ and $ \rho_{ AB_{\II}}^{\Phi_{*}}$
coincides at the infinite acceleration. }
\end{center}
\end{figure}

\subsection{Bipartite pure states III - particle and antiparticle degrees of freedom}
\label{sub3}

We now consider pure entangled states that describe correlations
between particle and anti-particle degrees of freedom, when Bob is
traveling with a uniform acceleration, as follows:
\begin{equation}
|\Phi_{*} (\alpha) \rangle = \cos \alpha |1\rangle^{+}_{M}|1^{+}\rangle_{U} + \sin
\alpha |1\rangle^{-}_{M}|1^{-}\rangle_{U}. \label{eq8}
\end{equation}
As was done before, the state that Alice and Bob share can be
obtained beyond the single-mode approximation. The state when Bob is
in region $\I$ is obtained by tracing the other region,
\begin{eqnarray}
\rho_{AB_{\mathrm{I}}}^{\Phi_{*}}
&=&  q_{L}^{2} \cos ^2 \alpha \cos ^2 \gamma |+00\rangle \langle +00| \nonumber\\
&+& \frac{1}{2}(1+(1-2 q_{L}^{2})\cos 2\gamma) \cos ^{2} \alpha (|+10\rangle \langle +10| ) \nonumber\\
&+& \frac{1}{4}(1+(1-2 q_{L}^{2})\cos 2\gamma) \sin 2 \alpha  |+10\rangle \langle -01|  \nonumber\\
&+& \frac{1}{4}(1+(1-2 q_{L}^{2})\cos 2\gamma) \sin 2 \alpha  |-01\rangle \langle +10|  \nonumber\\
&+& \frac{1}{2}(1+(1-2 q_{L}^{2})\cos 2\gamma) \sin ^{2}\alpha
|-01\rangle \langle -01| \nonumber\\
&+& q_{L}^{2} \sin ^{2} \alpha \cos ^{2} \gamma |-00\rangle \langle -00| \nonumber\\
&-& \frac{q_{R}q_{L}}{2} \cos ^{2}\alpha \sin 2\gamma (|+00\rangle
\langle +11| + |+11\rangle \langle +00| ) \nonumber\\
&-& \frac{q_{R}q_{L}}{2} \sin ^{2}\alpha \sin 2\gamma (|-00\rangle
\langle -11| + |-11\rangle \langle -00| ) \nonumber\\
&+& q_{R}^{2} \cos ^{2} \alpha \sin ^{2} \gamma |+11\rangle \langle +11| \nonumber\\
&+& q_{R}^{2} \sin ^{2} \alpha \sin ^{2} \gamma |-11\rangle \langle -11|. \nonumber
\end{eqnarray}
Also, the state that Alice and anti-Bob (in Bob's region II) share is,
\begin{eqnarray}
\rho_{AB_{\mathrm{II}}}^{\Phi_{*}}
&=&  q_{R}^{2} \cos ^2 \alpha \cos ^2 \gamma |+00\rangle \langle +00| \nonumber\\
&+& \frac{1}{2}(1+(1-2 q_{R}^{2})\cos 2\gamma) \cos ^{2} \alpha (|+10\rangle \langle +10| ) \nonumber\\
&+& \frac{1}{4}(1+(1-2 q_{R}^{2})\cos 2\gamma) \sin 2 \alpha   |+10\rangle \langle -01| \nonumber\\
&+& \frac{1}{4}(1+(1-2 q_{R}^{2})\cos 2\gamma) \sin 2 \alpha   |-01\rangle \langle +10|  \nonumber\\
&+& \frac{1}{2}(1+(1-2 q_{R}^{2})\cos 2\gamma) \sin ^{2}\alpha
|-01\rangle \langle -01| \nonumber\\
&+& q_{R}^{2} \sin ^{2} \alpha \cos ^{2} \gamma |-00\rangle \langle -00| \nonumber\\
&-& \frac{q_{R}q_{L}}{2} \cos ^{2}\alpha \sin 2\gamma (|+00\rangle
\langle +11| + |+11\rangle \langle +00| ) \nonumber\\
&-& \frac{q_{R}q_{L}}{2} \sin ^{2}\alpha \sin 2\gamma (|-00\rangle
\langle -11| + |-11\rangle \langle -00| ) \nonumber\\
&+& q_{L}^{2} \cos ^{2} \alpha \sin ^{2} \gamma |+11\rangle \langle +11| \nonumber\\
&+& q_{L}^{2} \sin ^{2} \alpha \sin ^{2} \gamma |-11\rangle \langle -11|. \nonumber
\end{eqnarray}
For these two states, entanglement is computed in terms of
negativity, and is shown in Fig. \ref{fig6}.

Interestingly, compared to other cases in Sec. \ref{sub1} and Sec.
\ref{sub2}(see Figs. \ref{fig2} and \ref{fig4}), the behavior
depicted in Fig. \ref{fig6} is shown to be nearly equidistant
according to different values of $q_{R}$. It is also shown that the
entanglement behavior of two states $ \rho_{ AB_{\I}}^{\Phi_{*}}$
and $ \rho_{ AB_{\II}}^{\Phi_{*}}$ is independent of $q_{R}$ at the
infinite acceleration, $\gamma=\frac{\pi}{4}$.

Let us also mention about when Bob and anti-Bob detector can
distinguish between the particle and the antiparticle. In this case,
it turns out that all density matrices corresponding to cases of
Alice-Bob particle in region I, Alice-Bob antiparticle in region I,
Alice-anti-Bob particle in region II, and Alice-anti-Bob
antiparticle in region II are separable. Thus, negativity remains
zero for all parameters.

\subsection{Bipartite mixed states - particle and antiparticle Unruh excitations}
\label{sub4}

Up to now, we have considered entanglement of pure states when one
of parties is traveling with a uniform acceleration. We have
observed that, at the infinite acceleration, entanglement has the
same convergence, which allows us to have physical results. In this
subsection, we consider a more complicate scenario when two parties
share a mixed state. It is our aim to discover how the entanglement
behavior depends on the mixedness property, and also if its
convergence is related to the mixedness. In particular, the case in
which a white noise is added to a maximally entangled states,
so-called Werner state, is to be considered. For Werner states, the
mixedness is parameterized by a single parameter, depending on which
one can see how noisy a state is.

Suppose that two parties, Alice and Bob prepare Werner states in
inertial frames, and then Bob moves in the uniformly accelerated
frame. That is, the state can be expressed as follows:
\begin{equation}
\rho_{W}= F |\Phi_{+} (\alpha=\pi/4) \rangle  \langle \Phi_{+} (\alpha=\pi/4) | + \frac{1-F}{4}\mathbb{I},
\end{equation}
where the maximally entangled state is taken from Eq. (\ref{eq6})
when $\alpha=\pi/4$.

Suppose also that Bob's detector cannot distinguish between the
particle or the antiparticle. Beyond the single-mode approximation,
both states that Alice and Bob share in Bob's region I and region
II, respectively, are obtained by tracing the other region, as
follows:

\begin{widetext}
\begin{eqnarray}
\rho_{AB_{\mathrm{I}}}^{W} &=& \frac{1}{2}F q_{R} \cos ^3 \gamma (|000\rangle \langle 110|+|110\rangle \langle 000|) + \frac{1}{8} \cos ^{2} \gamma (3-2 q_{R}^{2}+F(1-2q_{R}^{2})+(1-F)\cos 2\gamma)|100\rangle \langle 100| \nonumber\\
&+ & \frac{1}{8} \cos ^{2} \gamma (3-2 q_{R}^{2}-F(1-2q_{R}^{2})+(1+F)\cos 2\gamma)|000\rangle \langle 000| - \frac{F q_{L}}{2} \cos ^{2}\gamma \sin \gamma(|001\rangle \langle 100| + |100\rangle \langle 001|) \nonumber\\
&+&\frac{F q_{R}}{2} \cos \gamma \sin ^{2}\gamma(|001\rangle \langle 111| + |111\rangle \langle 001|) +\frac{F q_{L}}{2} \sin ^{3}\gamma(|011\rangle \langle 110| + |110\rangle \langle 011|) \nonumber\\
&+& \frac{1}{4} \sin ^{2} \gamma ((1+F)q_{R}^{2}+(1-F) \sin ^{2}\gamma)|111\rangle \langle 111| + \frac{1}{4} \sin ^{2} \gamma ((1-F)q_{R}^{2}+(1+F) \sin ^{2}\gamma)|011\rangle \langle 011| \nonumber\\
&-& \frac{1}{8} (1-F)q_{L}q_{R} \sin 2 \gamma (|000\rangle \langle 011| +|011\rangle \langle 000|) -  \frac{1}{8} (1+F)q_{L}q_{R} \sin 2 \gamma (|100\rangle \langle 111| +|111\rangle \langle 100|)\nonumber\\
&+&\frac{1}{16} (1-F) \sin ^{2} 2\gamma|101\rangle \langle 101|+\frac{1}{16} (1+F) \sin ^{2} 2\gamma|001\rangle \langle 001| + \frac{1}{16} (2(1+F)-2(1+F)(1-2 q_{R}^{2}) \cos 2\gamma \nonumber\\
& &+(1-F)\sin ^{2} 2\gamma)|110\rangle \langle 110| + \frac{1}{16} (2(1-F)-2(1-F)(1-2 q_{R}^{2}) \cos 2\gamma +(1+F)\sin ^{2} 2\gamma) |010\rangle \langle 010|, \nonumber
\end{eqnarray}
\end{widetext}
and
\begin{widetext}
\begin{eqnarray}
\rho_{AB_{\mathrm{II}}}^{W} & = & \frac{1}{2}F q_{L} \cos ^3 \gamma (|000\rangle \langle 110|+|110\rangle \langle 000|) + \frac{1}{8} \sin ^{2} \gamma (3-2 q_{R}^{2}+F(1-2q_{R}^{2}) - (1-F) \cos 2\gamma)|111\rangle \langle 111| \nonumber\\
&+& \frac{1}{8} \sin ^{2} \gamma (3-2 q_{R}^{2}-F(1-2q_{R}^{2}) -(1+F) \cos 2\gamma)|011\rangle \langle 011| + \frac{F q_{R}}{2} \cos ^{2}\gamma \sin \gamma(|001\rangle \langle 100| + |100\rangle \langle 001|) \nonumber\\
&-&\frac{F q_{L}}{2} \cos \gamma \sin ^{2}\gamma(|001\rangle \langle 111| + |111\rangle \langle 001|) + \frac{F q_{R}}{2} \sin ^{3}\gamma(|011\rangle \langle 110| + |110\rangle \langle 011|) \nonumber\\
&+& \frac{1}{4} \cos ^{2} \gamma ((1+F)q_{R}^{2}+(1-F) \cos ^{2}\gamma)|100\rangle \langle 100| +  \frac{1}{4} \cos ^{2} \gamma ((1-F)q_{R}^{2}+(1+F) \cos ^{2}\gamma)|000\rangle \langle 000| \nonumber\\
&-& \frac{1}{8} (1-F)q_{L}q_{R} \sin 2 \gamma (|000\rangle \langle 011| +|011\rangle \langle 000|) -  \frac{1}{8} (1+F)q_{L}q_{R} \sin 2 \gamma (|100\rangle \langle 111| +|111\rangle \langle 100|)\nonumber\\
&+&\frac{1}{16} (1-F) \sin ^{2} 2\gamma|101\rangle \langle 101| + \frac{1}{16} (1+F) \sin ^{2} 2\gamma|001\rangle \langle 001| + \frac{1}{16} (2(1+F)+2(1+F)(1-2 q_{R}^{2}) \cos 2\gamma \nonumber\\
& +&(1-F) \sin ^{2} 2\gamma)|110\rangle \langle 110| + \frac{1}{16} (2(1-F)+2(1-F)(1-2 q_{R}^{2}) \cos 2\gamma +(1+F) \sin ^{2} 2\gamma)|010\rangle \langle 010|. \nonumber
\end{eqnarray}
\end{widetext}

\begin{figure}
\begin{center}  
 \includegraphics[width=8cm]{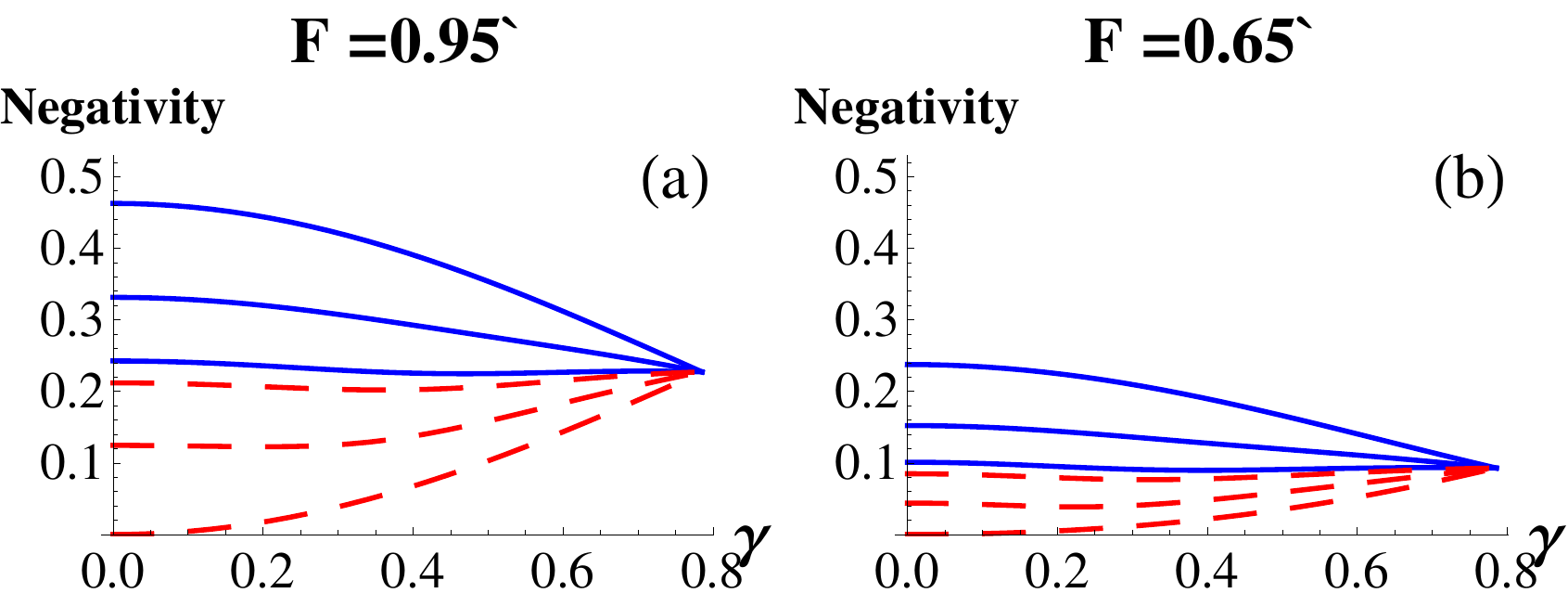}
  \caption{\label{fig7}(Color online) Entanglement
 of states $ \rho_{AB_{\mathrm{I}}}^{W}$ and $ \rho_{
AB_{\mathrm{II}}}^{W}$ is shown. Parts (a) and (b) show the cases of
$F=0.95$ and $F=0.65$ respectively. The blue solid(the red dotted)
lines from top to bottom(from bottom to top) denote the negativity
of  $ \rho_{AB_{\mathrm{I}}}^{W}$ ($ \rho_{ AB_{\mathrm{II}}}^{W}$)
at $q_{R}=1$,$q_{R}=0.85$ and $q_{R}=0.73$ respectively. At the
infinite acceleration, i.e. $\gamma=\frac{\pi}{4}$, it is shown that
the entanglement of $ \rho_{AB_{\mathrm{I}}}^{W}$ and $
\rho_{AB_{\mathrm{II}}}^{W}$ coincides.}
\end{center}
\end{figure}

Entanglement  of these states $\rho_{AB_{\mathrm{I}}}^{W}$ and $
\rho_{AB_{\mathrm{II}}}^{W}$ are shown in terms of negativity in
Fig. \ref{fig7}. It is shown that the entanglement behavior of
$\rho_{AB_{\mathrm{I}}}^{W}$ and $\rho_{AB_{\mathrm{II}}}^{W}$
coincides at, and is independent of $q_{R}$, at the infinite
acceleration $\gamma=\frac{\pi}{4}$.

Next, when Bob and anti-Bob detector can distinguish between the
particle and the antiparticle, density matrices can be found, for
the following cases, Alice-Bob particle in region I, Alice-Bob
antiparticle in region I, Alice-anti-Bob particle in region II, and
Alice-anti-Bob antiparticle in region II. From the matrices, it is
straightforward to compute entanglement using negativity. In Fig. 8,
the entanglement behavior is shown.

\begin{figure}[h!]
\begin{center}  
  \includegraphics[width=7.5cm]{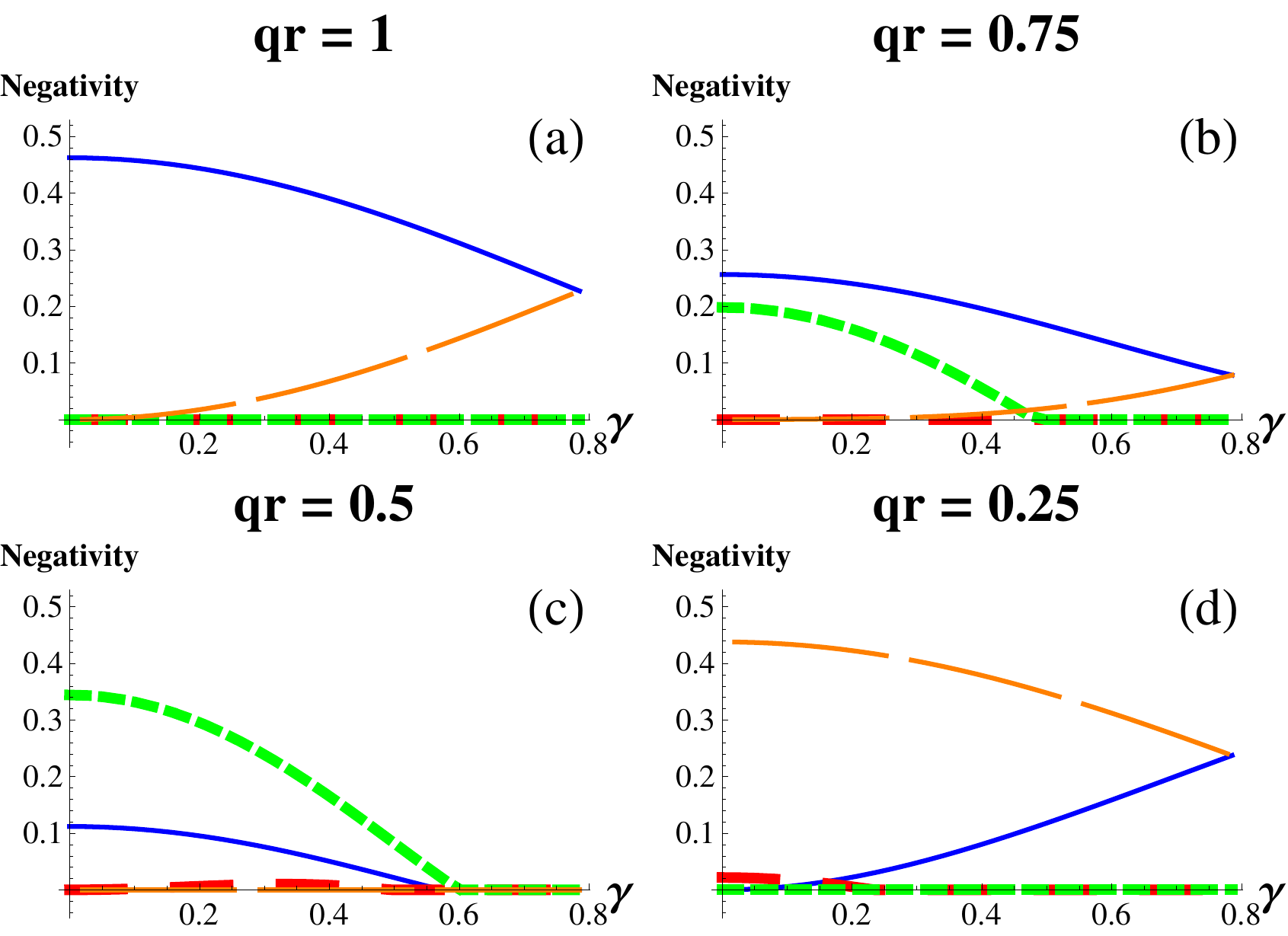}
  \caption{\label{fig8}(Color online) Entanglement of
 Alice-Bob particle in region I, Alice-Bob antiparticle in region
I, Alice-anti-Bob particle in region II, and Alice-anti-Bob
antiparticle in region II, for the state $\rho_{W}$ is shown. Parts
(a), (b), (c) and (d) show the cases of
$q_{R}=1,q_{R}=0.75,q_{R}=0.5$,and $q_{R}=0.25$ respectively (here
we set $F=0.95$). The blue solid line, the red thick dashed one, the
green dotted one and the orange dot-dashed one denote the negativity
of Alice-Bob particle in region I, Alice-Bob antiparticle in region
I, Alice-anti-Bob particle in region II, and Alice-anti-Bob
antiparticle in region II respectively.}
\end{center}
\end{figure}

\section{Discussion and Conclusion}
\label{con}

In this article, we have investigated the entanglement behavior of
bipartite quantum states in fermionic systems when one of parties is
traveling with a uniform acceleration. We have employed the recent
proposal in Ref. \cite{ref:montero1} for the ordering of operators.
This is because the ordering is suggested such that field
entanglement is relevant to what is observed in detectors. We
believe that this is a natural and relevant constraint, as it is
designed to yield physical results, which is contrary to other
previous approaches that consider various possibilities depending on
mathematical ordering. Before the present consideration, the
construction in Ref. \cite{ref:montero1} was only tested for a
particular pure state in Ref. \cite{ref:montero2}.

We have applied the construction and considered entanglement
behavior for numerous cases,i.e, bipartite entanglement of pure and
mixed states. We have shown that, in all of these cases, the
entanglement behavior allows one to obtain physical results so that
in the infinite acceleration, entanglement converges to a single and
finite value. Our considerations consist of exemplary states shown
in Ref. \cite{ref:martin}, where a different ordering has been
applied and consequently the convergence property has not been
achieved in the entanglement behavior. This contrasts to what is
shown in the present work, and thus our result has provided the
correct behavior of entanglement in fermonic systems.

\section*{Acknowledgement}
 We would like to thank Dr.Mart\'{i}n-Mart\'{i}nez and Dr.Joonwoo Bae for careful reading of the manuscript and valuable comments on it.This work is supported by Basic Science Research Program through the National Research Foundation of Korea funded by the Ministry of Education, Science and Technology (KRF2010-0025620 and KRF2011-0027142).

\end{document}